\documentclass[default]{svmult}
\input{epsf}

\usepackage{mathptmx}       
\usepackage{helvet}         
\usepackage{courier}        
\usepackage{type1cm}        
\usepackage{makeidx}         
\usepackage{graphicx}        
\usepackage{multicol}        
\usepackage[bottom]{footmisc}

\makeindex             

\newcommand{\apj}{ApJ}

\newcommand{\aj}{AJ}
\newcommand{\mnras}{MNRAS}

\newcommand{\aap}{A\&A}

\begin{document}

\title*{Extremely Red Objects in a hierarchical universe}
\author{
V.\,Gonzalez-Perez,
C.\,M.\,Baugh,
C.G.Lacey,
C. Almeida}
\institute{V.\,Gonzalez-Perez \at Institut de Ci\`{e}ncies de l'Espai (CSIC/IEEC), F. de Ciencies, Torre C5 Par 2a, UAB, Bellaterra, 08193 Barcelona, Spain, \email{gonzalez@ieec.uab.es}
\and C.\,M.\,Baugh, C.G.Lacey, C. Almeida \at Institute for Computational Cosmology, Department of Physics, 
University of Durham, South Road, Durham, DH1 3LE, UK
}
%
%
\maketitle

\abstract*{}

\abstract{We analyse whether hierarchical formation models based on $\Lambda$ cold dark matter cosmology can produce enough massive red galaxies to match observations. For this purpose, we compare with observations the predictions from two published models for the abundance and redshift distribution of Extremely Red Objects (EROs), which are red, massive galaxies observed at $z \geq 1$. One of the models invokes a ``superwind'' to regulate star formation in massive haloes and the other suppresses cooling through ``radio-mode'' AGN feedback. The first one underestimates the number counts of EROs by an order of magnitude, whereas the radio-mode AGN feedback model gives excellent agreement with the number counts of EROs and redshift distribution of K-selected galaxies. This study highlights the need to consider AGN feedback in order to understand the formation and evolution of massive galaxies at $z \geq 1$.}

\section{Introduction}
\label{sec:intro}
In hierarchical models, dark matter structures grow hierarchically, i.e., more massive haloes form most recently. However, this does not necessarily imply
that the most massive galaxies formed the most recently. If the formation of galaxy mass tracked that of the dark halo, then there would be a problem for the formation of massive galaxies seen at $z>1$. Extremely Red Objects (EROs) allows us to study this point since they are red - $(R-K)>5$ , massive galaxies, observed at $<z>\sim 1.1$ \cite{cimatti03} (this value depends on the magnitude limit of the observational survey). The extremely red optical-near infrared colours of EROs, indicate either a starforming galaxy 
with heavy obscuration or an aged stellar population with little recent 
star formation \cite{smail02}. In the latter case, the implied age of the stellar population is uncomfortably close to the age of the Universe at the redshift of observation \cite{smith02}. 

Reproducing the abundance of EROs has previously eluded many hierarchical models. Recently there has been much development in the modelling of the formation of massive galaxies. Typically, hierarchical models produce too many massive galaxies at the present day unless some physical process is invoked to restrict gas cooling in massive haloes (see e.g. \cite{benson03} for a review of plausible mechanisms). The most promising candidates are the heating of gas in massive haloes by ``AGN feedback'' \cite{bower06} and the ejection of gas from intermediate mass haloes in a wind, which could be driven by supernovae \cite{baugh05} or by a QSO phase of accretion onto a black hole \cite{granato04}. 

The bands used here are centred at $0.65$ $\mu m$, R band, and  $2.2$ $\mu m$, K band. All magnitudes are on the Vega system, unless otherwise specified. 

The results described in this proceedings are taken from
Gonzalez-Perez (2008) \cite{eros1}. 

\section{Galaxy formation models}

We study the abundance and properties of EROs in a $\Lambda$CDM universe using the semi-analytical galaxy formation code {\sc galform} \cite{cole00}. Semi-analytical models apply simple physical recipes to determine the fate of the baryonic component of the universe. 

For this work we focus on two published developments of {\sc galform} \cite{cole00}: the Baugh et al.\cite{baugh05} and Bower et al.\cite{bower06} models. We extract predictions for EROs galaxies without adjusting the values of any of the model parameters. These parameters were fixed with reference to a subset of observational data, mostly at low redshift. Both models have some notable successes at higher redshifts: the Baugh et al.\cite{baugh05} model reproduces the number counts of sub-mm selected galaxies and the luminosity function of Lyman-break galaxies, whereas Bower et al.\cite{bower06} model matches the evolution of the K-band luminosity function and the inferred evolution of the stellar mass function.

\begin{table}
\caption{Basic differences between the two models used in this work:}
\label{tab:mod}       
\begin{tabular}{ccc}
  & {\bf Baugh et al. 2005} & {\bf Bower et al. 2006}  \\ \hline
$\Omega_{0}$ & $0.3$ & $0.25$ \\
$\Lambda_{0}$ & $0.7$ &$0.75$ \\
$\Omega_{b}$ & $0.04$ & $0.045$\\
$\sigma_{8}$ & $0.93$ & $0.90$ \\
$h$ & $0.7$ & $0.73$\\\hline
Dark matter halo merger trees   & Monte Carlo & N-body \\ \hline
Quenching of star formation in bright galaxies & Superwind & AGN feedback\\ \hline
Time scale for quiescent star formation & Independent of time  & Dependent on time\\ \hline
Bursts triggered by   & Mergers & Mergers and disk instabilities \\ \hline
Burst Initial Mass Function    & Top heavy &  Solar neighbourhood\cite{kennicutt_imf}  \\
\hline
\end{tabular}
\end{table}

 Table \ref{tab:mod} lists  the cosmological parameters used in each model and the main differences between them. For our study, the key difference is the physics used to suppress star formation within bright galaxies: Baugh et al.\cite{baugh05} introduce an outflow wind that ejects cooled gas from haloes, while in Bower et al.\cite{bower06} model AGN feedback prevents gas from cooling in massive enough galaxies.

\section{Results}

\begin{figure}[b]
\includegraphics[scale=.55]{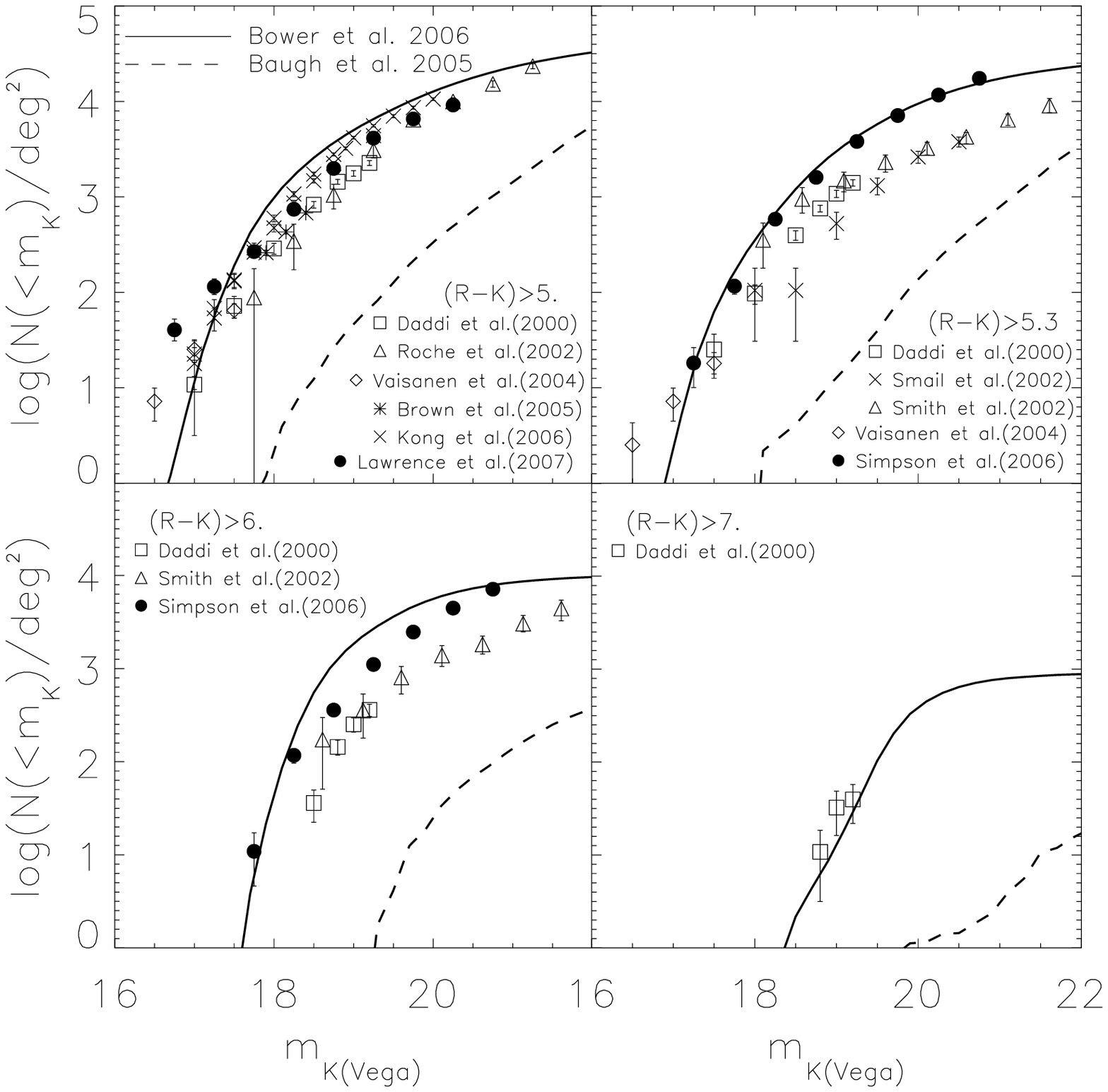}
%
%
\caption{K-band cumulative number counts for EROs selected by their $(R-K)$
  colour: $(R-K)>5$ in top left panel, $(R-K)>5.3$ in top
  right panel, $(R-K)>6$ in bottom left panel and $(R-K)>7$ in the
  bottom right panel. The solid lines corresponds to the predictions
  from Bower et al.\cite{bower06} and the dashed ones to the Baugh et al.\cite{baugh05} model. The source of the observational 
  data is given in the legend \cite{daddi00,roche02,smith02,smail02,vaisanen04,brown05,simpson06,kong06,lawrence07} of each graph and it can be find at: http://segre.ieec.uab.es/violeta/rawEROSdata.txt. The errors shown are Poisson.
}
\label{fig:ncounts}       
\end{figure}

\begin{figure}[h]
{\epsfxsize=12truecm
\epsfbox[50 27 490 385]{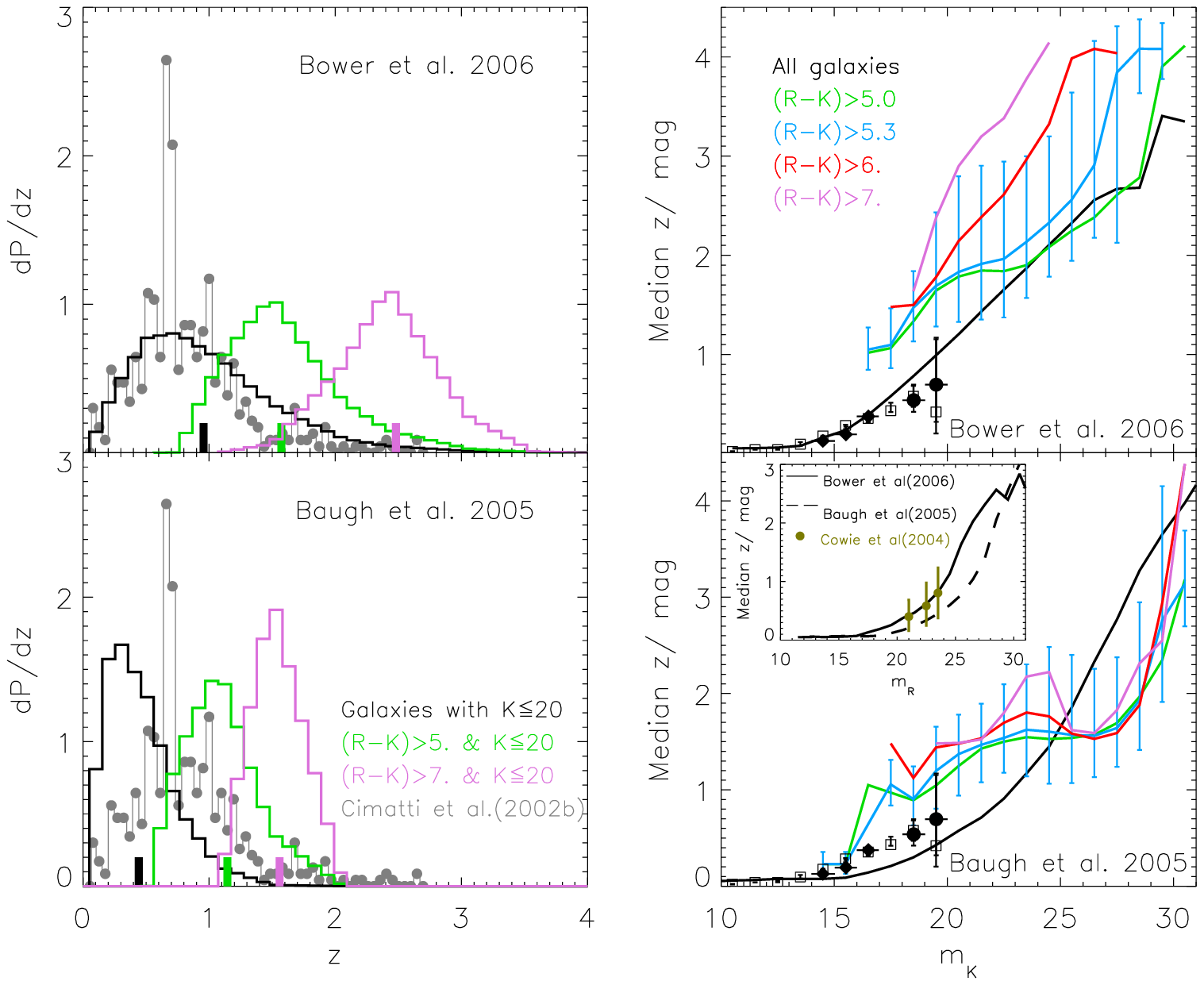}}
\caption{{\it Left} The redshift distribution of EROs with $K\leq 20$, defined by $(R-K)$ colour redder
 than $5$ (green histogram) and $7$ (pink). The black histogram represents 
the redshift distribution for all galaxies with $K\leq 20$. The marks on the x-axis show the median of the distribution plotted with the same colour. 
Observational data \cite{cimatti02b}
for galaxies with $K<20$ is shown as grey points. 
The upper panel shows the predictions from Bower et al.\cite{bower06} and the lower one 
from the Baugh et al.\cite{baugh05} model. Histograms are normalised to give unit area under the distribution.
{\it Right:} The median redshift as a function of K band magnitude for all galaxies (black) and
for EROs with $(R-K)$ colour redder than  $5$ (green), $5.3$ (blue), $6$ (red) and 
$7$ (pink). For clarity, quartiles are only shown for EROs with $(R-K)>5.3$. 
The upper panel shows predictions from Bower et al.\cite{bower06} and the lower one 
from Baugh et al.\cite{baugh05}. For comparison we plot the median redshift for all galaxies 
per magnitude bin in the $K$-band observed in different studies (squares\cite{songaila90}, 
diamonds \cite{glazebrook95} and circles \cite{cowie96}). The inset shows the 
median redshift per magnitude bin in the $R$-band for all galaxies predicted 
by Bower et al.\cite{bower06} (solid line) and Baugh et al.\cite{baugh05} (dashed line), and 
observational \cite{cowie04} as circles.
}
\label{fig:z}       
\end{figure}

In Fig. \ref{fig:ncounts} we compare the observed EROs number counts with the predicted ones, for different colour cuts defining the ERO population. The predictions of the Bower et al.\cite{bower06} model reproduce the observed abundance of EROs impressively well for all the colour cuts shown, which is remarkable as none of the model parameters have been tuned to achieve this level of agreement. 

The agreement is less impressive for the Baugh et al.\cite{baugh05} model. Fig. \ref{fig:ncounts} shows that this model typically underestimates the number counts of EROs by more than an order of magnitude. This lack of EROs cannot be blamed on a fluctuation in the observed counts due to sampling variations arising from the small fields and strong clustering of EROs.

Some fraction of EROs are likely to be heavily dust extincted. It is important therefore to make a realistic calculation of the degree to which galaxy colours are extincted. We have experimented with changing and improving the dust treatment, without finding any significant change in the predicted ERO counts. We also found out that shortening the burst duration in Baugh et al.\cite{baugh05} model had little impact on the predicted number of EROs. The small number of dusty EROs predicted by this model is remarkable,
and this could in part explain the under-prediction of the total ERO
counts. Therefore, the disagreement between Baugh et al.\cite{baugh05} model and observations suggests that the superwind feedback used in this model could be too efficient for bright galaxies at $z>1$, delaying their formation.

The predicted differential redshift distribution of galaxies brighter than $K=20$ and EROs selected with different colour cuts are shown in the left panel of Fig. \ref{fig:z}. It can be seen that Bower et al.\cite{bower06} makes an accurate prediction of the redshift distribution of K-selected galaxies up to $z\sim 1.3$. Such a good agreement is in contrast with the Baugh et al.\cite{baugh05} model which predicts a tighter redshift distribution around a lower redshift value. 

Fig. \ref{fig:z} shows that the observations of significant numbers of EROs with $z>1.5$ favours the Bower et al.\cite{bower06} model; in the Baugh et al.\cite{baugh05} model it is much less likely that one would find galaxies at such redshifts.

Within the {\sc ukidss} survey data \cite{simpson06} a clear tendency was seen for redder EROs to be at higher redshifts. The same tendency is found for both models, as seen in the left panel of Fig. \ref{fig:z}.

For further comparison, we plot in the right panel of Fig. \ref{fig:z} the observed median redshift distribution for both the K and R bands and compare with the models predictions. The Bower et al.\cite{bower06} prediction is consistent with  the observations in both bands. The Baugh et al.\cite{baugh05} model matches the data better in the R-band than it does in the K-band. This model underestimates the median redshift for K-selected samples faintwards of $K\sim 16$.

\section{Conclusions}

In this work we have extended the tests of the {\sc galform} galaxy formation code to include galaxies at $z>1$. We have tested two published models of galaxy formation, those of Baugh et al.\cite{baugh05} and Bower et al.\cite{bower06}, against observed EROs number counts and K-selected redshift distributions. 

The Bower et al.\cite{bower06} model gives an impressively close match to the number counts of EROs, while Baugh et al.\cite{baugh05} predicts an order of magnitude too few EROs. Also, Baugh et al.\cite{baugh05} model predicts a too low median redshift for K-selected galaxies. We have experimented with Baugh et al.\cite{baugh05} model without finding a way to reconcile its predictions with observations. The key seems to be that Bower et al.\cite{bower06} model gives a better match to the observed evolution of the K-band luminosity function, which means that this model puts massive galaxies in place earlier than in the Baugh et al.\cite{baugh05} model. This difference between the two models arises from the different redshift dependence of the feedback processes which suppress the formation of massive galaxies and from the choice of the star formation time-scale.

\small{\section*{Acknowledgements}
We thank S. Foucaud, C. Simpson, I. Smail and F. J. Castander. VGP acknowledges support from CSIC/IEEC, the Spanish Ministerio de Ciencia y Tecnolog\'{i}a and travel support from a Royal Society International Joint Project grant. CMB is supported by the Royal Society. CGL is supported in part by a grant from the Science and Technology Facilities Council. CA gratefully acknowledges a scholarship from the FCT, Portugal.
}

%
%
%

\end{document}